\documentclass[sigconf]{acmart}

\AtBeginDocument{%
  \providecommand\BibTeX{{%
    \normalfont B\kern-0.5em{\scshape i\kern-0.25em b}\kern-0.8em\TeX}}}

\setcopyright{acmlicensed}
\copyrightyear{2018}
\acmYear{2018}
\acmDOI{XXXXXXX.XXXXXXX}

\acmConference[ICSE]{Make sure to enter the correct
  conference title from your rights confirmation emai}{2024, April 2024}{Lisbon, Portugal}
\acmISBN{978-1-4503-XXXX-X/18/06}
\renewcommand\footnotetextcopyrightpermission[1]{}
\usepackage{multirow}
\usepackage[ruled,linesnumbered]{algorithm2e}
\usepackage{subcaption}
\usepackage{color,xcolor,colortbl}
\usepackage{enumitem}
\usepackage{graphicx}
\usepackage{tcolorbox}
\usepackage{bbding}
\usepackage{microtype}

\newcommand{\dataset}{ReposVul\xspace}
\newcommand{\moduleA}{vulnerability untangling module\xspace}
\newcommand{\moduleB}{multi-granularity dependency extraction module\xspace}
\newcommand{\moduleC}{trace-based filtering module\xspace}
\newcommand{\moduleCa}{file path trace-based filter\xspace}
\newcommand{\moduleCb}{commit time trace-based filter\xspace}

\newcommand{\fileA}{vulnerability-fixing related files\xspace}
\newcommand{\fileB}{vulnerability-fixing unrelated files\xspace}
\newcommand{\ie}{\textit{i}.\textit{e}.\xspace}

\begin{document}

\title{\dataset: A Repository-Level High-Quality Vulnerability Dataset}

\author{Xinchen Wang$^{\star}$}
\affiliation{%
  \institution{Harbin Institute of Technology,}
  \city{Shenzhen}
  \country{China}}
\email{200111115@stu.hit.edu.cn}

\author{Ruida Hu$^{\star}$}
\affiliation{%
  \institution{Harbin Institute of Technology,}
  \city{Shenzhen}
  \country{China}}
\email{200111107@stu.hit.edu.cn}

\author{Cuiyun Gao$^{\ast}$}
\affiliation{%
  \institution{Harbin Institute of Technology,}
  \city{Shenzhen}
  \country{China}}
\email{gaocuiyun@hit.edu.cn}

\author{Xin-Cheng Wen}
\affiliation{%
  \institution{Harbin Institute of Technology,}
  \city{Shenzhen}
  \country{China}}
\email{xiamenwxc@foxmail.com}

\author{Yujia Chen}
\affiliation{%
  \institution{Harbin Institute of Technology,}
  \city{Shenzhen}
  \country{China}}
\email{yujiachen@stu.hit.edu.cn}

\author{Qing Liao}
\affiliation{%
  \institution{Harbin Institute of Technology,}
  \city{Shenzhen}
  \country{China}}
\email{liaoqing@hit.edu.cn}

\thanks{$^{\star}$ These authors contribute to the work equally and are co-first authors of the paper.
 \\
$^{\ast}$ Corresponding author. The author is also affiliated with Peng Cheng Laboratory and Guangdong Provincial Key Laboratory of Novel Security Intelligence Technologies.
}

\begin{abstract}

\label{sec:abstract}

Open-Source Software (OSS) vulnerabilities bring great challenges to the software security and 
pose potential risks to our society. Enormous efforts have been devoted into automated vulnerability detection, among which deep learning (DL)-based approaches have proven to be the most effective. However, the performance of the DL-based approaches generally relies on the quantity and quality of labeled data, and the current labeled data present the following limitations:
(1) \textbf{Tangled Patches}: Developers may submit code changes unrelated to vulnerability fixes within patches, leading to tangled patches.
(2) \textbf{Lacking Inter-procedural Vulnerabilities}: The existing vulnerability datasets typically contain function-level and file-level vulnerabilities, ignoring the relations 
between functions, thus rendering the approaches unable to detect the inter-procedural vulnerabilities. (3) \textbf{Outdated Patches}: The existing datasets usually contain outdated patches, which may bias the model during training.

To address the above limitations, in this paper, we propose an automated data collection framework and construct the first repository-level high-quality vulnerability dataset named \textbf{\dataset}. The proposed framework 
mainly contains three modules: (1) A vulnerability untangling 
module, aiming at distinguishing vulnerability-fixing related code changes from tangled patches, in which the Large Language Models (LLMs) and static analysis tools are jointly employed. (2) A multi-granularity dependency extraction module, aiming at capturing the inter-procedural call relationships of vulnerabilities, in which we construct multiple-granularity information for each vulnerability patch, including repository-level, file-level, function-level, and line-level. (3) A trace-based filtering module, aiming at filtering the outdated patches, which leverages the file path trace-based filter and commit time trace-based filter to construct an up-to-date dataset.

The constructed repository-level \dataset encompasses 6,134 CVE entries representing 236 CWE types across 1,491 projects and four programming languages. Thorough data analysis and manual checking demonstrate that \dataset is high in quality
and alleviates the problems of tangled and outdated patches in previous vulnerability datasets.

\end{abstract}



\maketitle

\section{Introduction}

\label{sec:introduction}
\begin{figure*}[htb]
    \centering
    \includegraphics[width=\textwidth]{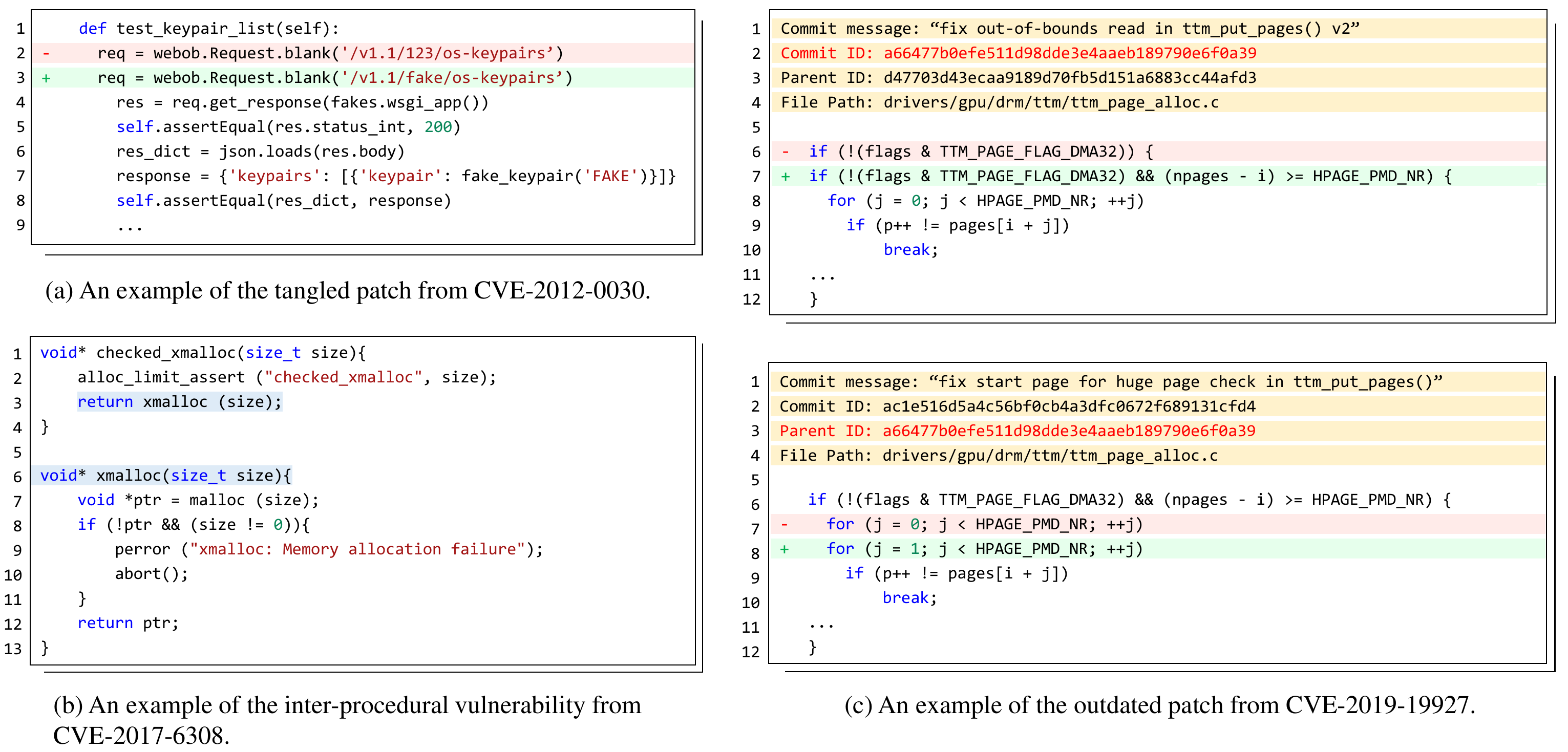}
    \caption{Examples for illustrating the challenges of existing datasets. 
    Lines highlighted in green denote added content, red indicates deleted content, yellow represents commit information, and blue identifies the caller and callee.
    }
    \label{fig:example}
\end{figure*}

In recent years, with the increasing size and complexity of Open-Source Software (OSS), the impact of OSS vulnerabilities has also amplified and can cause great losses to our society.
For example, Cisco discovered a security vulnerability in the WebUI, identified as CVE-2023-20198~\cite{cisco} in 2023. This vulnerability allowed unauthorized remote attackers to gain elevated privileges. Currently, over 41,000 related devices have been compromised, resulting in great losses for enterprises. Identifying vulnerabilities in an accurate and timely manner is beneficial for mitigating the potential risks, and has gained intense attention from industry and academia. The existing vulnerability detection methods can be coarsely grouped into two categories: program analysis-based methods~\cite{static1, static2, fuzz1, fuzz2} and deep learning (DL)-based methods~\cite{vul1,vul2,vul3,vul4}, among which DL-based methods have proven to be more effective. Despite the success of the DL-based methods, their performance tends to be limited by the trained vulnerability datasets. For example, Croft et al.~\cite{quality} find that the widely-used vulnerability datasets such as Devign~\cite{devign} and BigVul~\cite{bigvul} contain noisy, incomplete and outdated data. The low-quality data may bias the model training and evaluation process. Therefore, a high-quality real-world vulnerability dataset is important yet under-explored for the vulnerability detection task. In the paper, we focus on building the high-quality dataset by mitigating the following limitations of the existing datasets: 

\textbf{(1) Tangled Patches:} Vulnerability patches may contain vulnerability-fixing unrelated code changes, resulting in \textit{tangled patches}. Existing datasets~\cite{reveal, devign, bigvul} generally consider all the code changes in one patch submission related to the vulnerabilities, introducing natural data noise. For the example shown in Figure~\ref{fig:example}(a), the patch from CVE-2012-0030 includes the code change about path modification, in which the request path parameter of the function \texttt{webob.Request.blank()} has been changed (Lines 2-3). Such vulnerability-unrelated changes may be concerned with code refactoring or new feature implementation, and are hard to be distinguished. Therefore, identifying \fileA from multiple files in one patch presents a challenge. 


\textbf{(2) Lacking Inter-procedural Vulnerabilities:} Vulnerabilities in real-world scenarios usually involve calls between multiple files and functions, whereas individual functions alone are not necessarily vulnerable. Existing datasets~\cite{reveal, devign} mainly focus on function-level granularity, ignoring the call information. Figure~\ref{fig:example}(b) illustrates an example of inter-procedural vulnerability (CVE-2017-6308). The return value of the function \texttt{checked\_xmalloc()} is passed by the function \texttt{xmalloc()} (Line 3) in this example. Given that the parameter of \texttt{size} could overflow, there is a potential risk of triggering CWE-190 (Integer Overflow or Wraparound). However, the functions \texttt{checked\_xmalloc()} and \texttt{xmalloc()} are flawless themselves, while
existing labeling methods may mark them as vulnerable. Models trained on the vulnerable data without considering inter-procedural call information would be biased, limiting their performance in practical scenarios.

\textbf{(3) Outdated Patches:} 
Patches may introduce new vulnerabilities and become
outdated, while current datasets do not consider the timeliness of the patches.
Figure~\ref{fig:example}(c) shows an original
patch and its child patch both from CVE-2019-19927. In the original
patch, developers add stronger constraints to the existing conditional statement to avoid CWE-125 (Out-of-bounds Read)~\cite{CWE125} (Lines 6-7). However, in the next immediate patch, the loop statement after the conditional statement is changed (Lines 7-8), with the commit message stating ``fix start page for huge page check''. Clearly, this additional fix is due to the incompleteness of the original
patch, and thus the original patch is outdated and should be filtered out.




In this paper, we propose an automated data collection framework and construct a repository-level high-quality vulnerability dataset named \textbf{\dataset} to address the aforementioned limitations.
Our framework consists of three modules: \textcircled{1} \textbf{A \moduleA}: We propose to integrate the decisions of Large Language Models (LLMs) and static analysis tools to distinguish the \fileA within the patches, given their strong contextual understanding capability and domain knowledge, respectively. \textcircled{2} \textbf{A \moduleB}: We extract the inter-procedural call relationships of vulnerabilities among the whole repository, aiming to 
construct multi-granularity information for each vulnerability patch, including file-level, function-level, and line-level information. \textcircled{3} \textbf{A \moduleC}: We first track the submission history of patches based on file paths and commit time. Through analyzing historical information on the patches, we then identify outdated patches by tracing their commit diffs.

In summary, our contributions can be outlined as follows:
\begin{enumerate}

    \item 
We introduce an automated data collection framework for obtaining vulnerability data. Our framework consists of a \moduleA to identify \fileA within tangled patches, a \moduleB to construct inter-procedural vulnerabilities, and a \moduleC to recognize outdated patches.

    \item 
\dataset is the first repository-level vulnerability dataset, including large-scale CVE entries representing 236 CWE types across 1,491 projects and four programming languages with detailed multi-granularity patch information.

    \item 
Through manual checking and data analysis, \dataset is high in quality and alleviates the limitations of the existing vulnerability datasets. We have publicly released the source code as well as \dataset at: \url{https://github.com/Eshe0922/ReposVul}.


\end{enumerate}

The remaining sections of this paper are organized as follows. Section \ref{sec:data collection} introduces the framework to collect \dataset.
Section \ref{sec:evaluation} presents the evaluation and experimental results. Section \ref{sec:discussion} discusses the data application and limitations of \dataset. Section \ref{sec:related work} introduces the background of the OSS vulnerability datasets and detection methods. Section \ref{sec:conclustion} concludes the paper.

\section{Framework}

\label{sec:data collection}
\begin{figure*}[t]
	\centering
	\includegraphics[width=1\textwidth]{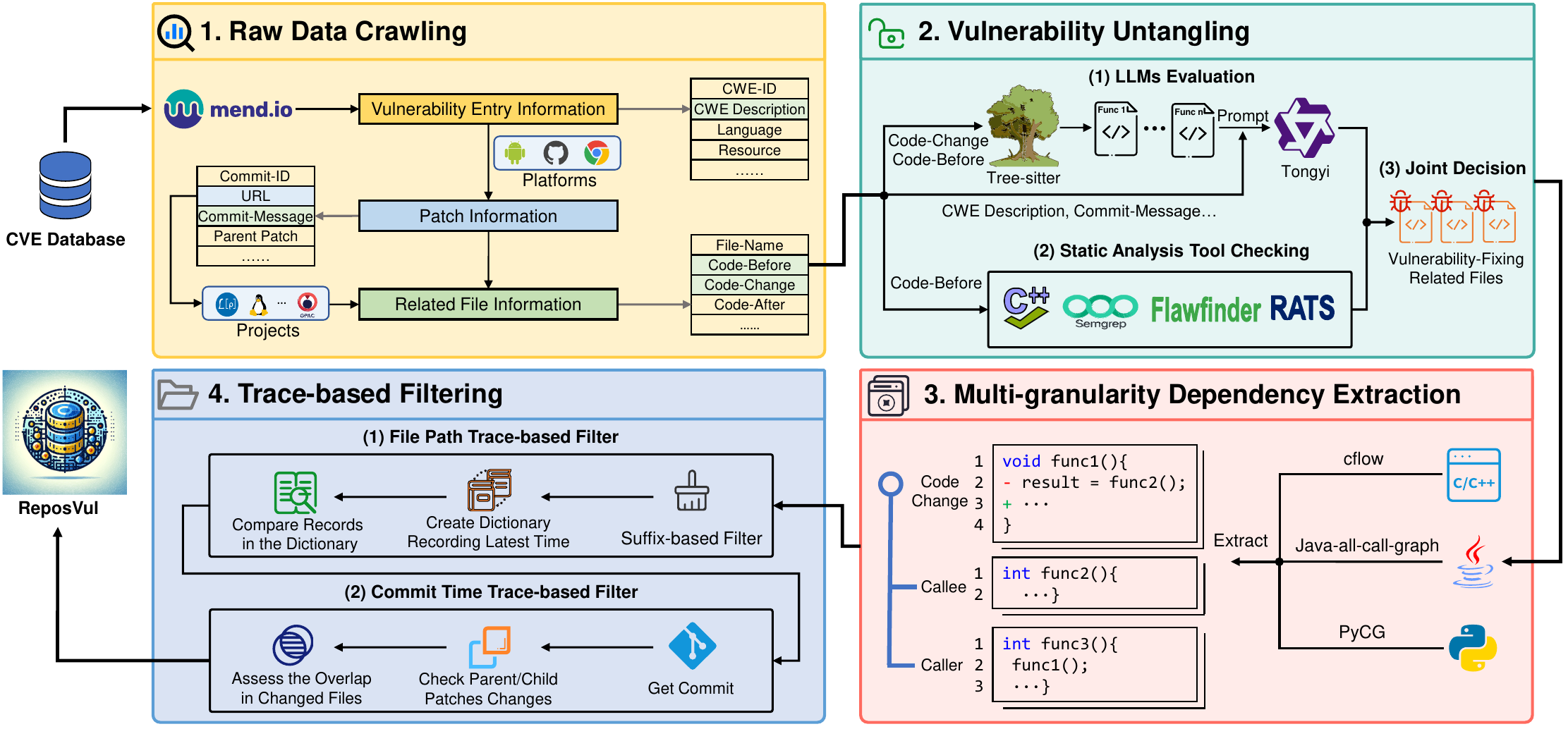}
    \caption{The architecture of our automatic data collection framework. 
    }
\label{fig:architecture}
\end{figure*}

Figure~\ref{fig:architecture}
presents an overview of our data collection framework, which contains four procedures
to construct the vulnerability dataset.
The framework begins with the \textit{raw data crawling} step, designed to gather vulnerability entries and associated
patches, resulting in an
initial dataset. This dataset is then processed through the following three key modules to yield the final dataset \dataset:
(1) The \textbf{\moduleA} aims at automatically identifying
\fileA within tangled patches, by jointly considering the decisions from LLMs and static analysis tools.
(2) The \textbf{\moduleB} extracts inter-procedural call relationships associated with vulnerabilities throughout the repository.
(3) The \textbf{\moduleC} tracks the submission history of patches based on file paths and commit time, aiming to analyze commit diffs corresponding to patches.

\subsection{Raw Data Crawling}
\begin{table}[t]
    \caption{Basic information for each entry in \dataset, including vulnerability entry information, patch information, and related file
    information.}
    \centering
    \setlength{\tabcolsep}{2mm}
    \renewcommand{\arraystretch}{1.0}
    \begin{tabular}{cl}
        \toprule
         \textbf{Features} &  \multicolumn{1}{l}{\textbf{Description}} \\
        \midrule
        \midrule
         \multicolumn{2}{c}{\textbf{Vulnerability Entry Information}} \\ \midrule
         CVE-ID &  The CVE that the entry belongs to \\
          CWE-ID & The CWE that the entry belongs to \\
          Language & The programming language of the CVE \\
          Resource & Involved links of the CVE\\
          CVE Description & The description of the CVE \\
          Publish-Date &  The publish date of the CVE\\
          CVSS &  The CVSS score of the CVE  \\
          CVE-AV &  The attack vector of the CVE  \\
          CVE-AC &  The attack complexity of the CVE  \\
          CVE-PR &  The privileges required of the CVE  \\
          CVE-UI &  The user interaction of the CVE  \\
          CVE-S &  The scope of the CVE  \\
          CVE-C &  The confidentiality of the CVE  \\
          CVE-I &  The integrity of the CVE  \\
          CVE-A &  The availability of the CVE  \\
          CWE Description & The description of the CWE \\
          CWE Solution & Potential solutions of the CWE\\
          CWE Consequence & Common consequences of the CWE\\
          CWE Method &  Detection methods of the CWE\\ 
          
        \midrule
        \midrule
        \multicolumn{2}{c}{\textbf{Patch Information}} \\ \midrule
          Commit-ID & The commit id of the patch\\
          Commit-Message &  The commit message of the patch \\ 
          Commit-Date &  The commit date of the patch \\ 
          Project & The project that the patch belongs to  \\  
          Parent Patch & The parent patch of the patch\\
          Child Patch & The child patch of the patch \\
          URL & The API-URL of the patch  \\ 
          Html-URL & The Html-URL of the patch  \\ 

        \midrule
        \midrule
        \multicolumn{2}{c}{\textbf{Related File
        Information}} \\ \midrule
          File-Name &  The name of the file \\ 
          File-Language &  The programming language of the file \\ 
           Code-Before &  The content of the file before fixes \\
          Code-After &  The content of the file after fixes\\ 
           Code-Change &  The code changes of the file \\
             Html-URL & The Html-URL of the file \\ 
        \bottomrule
    \end{tabular}
    \label{tab:dataset_info}
\end{table}

This phase aims to collect the extensive raw vulnerability data, with detailed entry information illustrated in Table~\ref{tab:dataset_info}. The creation of the initial dataset involves three steps: 1)
crawling vulnerability entries from open-source databases, 2)
fetching
patches associated with the vulnerability entry from multiple platforms, and 3)
obtaining detailed information on changed files involved in the patch.


\subsubsection{Vulnerability Entry Collection}

We collect open-source vulnerabilities from Mend~\cite{Mend}
which involves both popular and under-the-radar community resources with extensive vulnerability entries. During the collection, we retrieve
CVE entries in chronological order for the identification of outdated patches. We then store these entries in a structured format, as illustrated in the ``Vulnerability Entry Information'' part of Table~\ref{tab:dataset_info}. Each entry encompasses essential features such as the CVE-ID, CVE description, associated CWE-ID, and other relevant information. 


\subsubsection{Patch Collection}

To comprehensively analyze each vulnerability entry, we collect the corresponding patches. For the majority of the
projects, we
collect these patches from GitHub~\cite{GitHub} and record their Commit-ID and Commit-Message. Additionally, for two special projects, including Android and Chrome, we conduct patch collection on Google Git~\cite{GoogleGit} and bugs.chromium~\cite{bugs.chromium}, respectively, as some of their patches are not released
on GitHub. Detailed patch-level information
is summarized in Table~\ref{tab:dataset_info}.

\subsubsection{Related File Collection}

To extract vulnerability code snippets at file-level and repository-level, we download the entire repository of the parent patch and child patch associated with each patch 
using its unique Commit-ID.
For each 
file in the patch, we retrieve its content before and after code changes. 
In Table~\ref{tab:dataset_info}, we present the details of each related file, including key features like File-Name, Code-Before, and Code-After.



\label{sec:solution}
\subsection{Vulnerability Untangling Module}




The vulnerability untangling module aims to remove vulnerability-fixing unrelated code changes from patches.
We employ
LLMs for evaluating the relevance between
code changes and vulnerability fixes, and static analysis tools for checking vulnerabilities in the
code changes, separately. Jointly considering their outputs, we determine whether a changed file is vulnerability-fixing related.

\subsubsection{LLMs Evaluation}
LLMs possess strong contextual understanding capability and extensive
applicability across programming languages. The adaptability of LLMs to different programming languages also ensures that LLMs-based 
evaluation can be conducted across various code syntaxes and structures. We opt for Tongyi~\cite{TongYi} to evaluate the relevance between code changes and vulnerability fixes effectively, considering its accessible free API and support for long context inputs.

\begin{figure}[tb]
    \centering
    \includegraphics[width=0.48\textwidth]{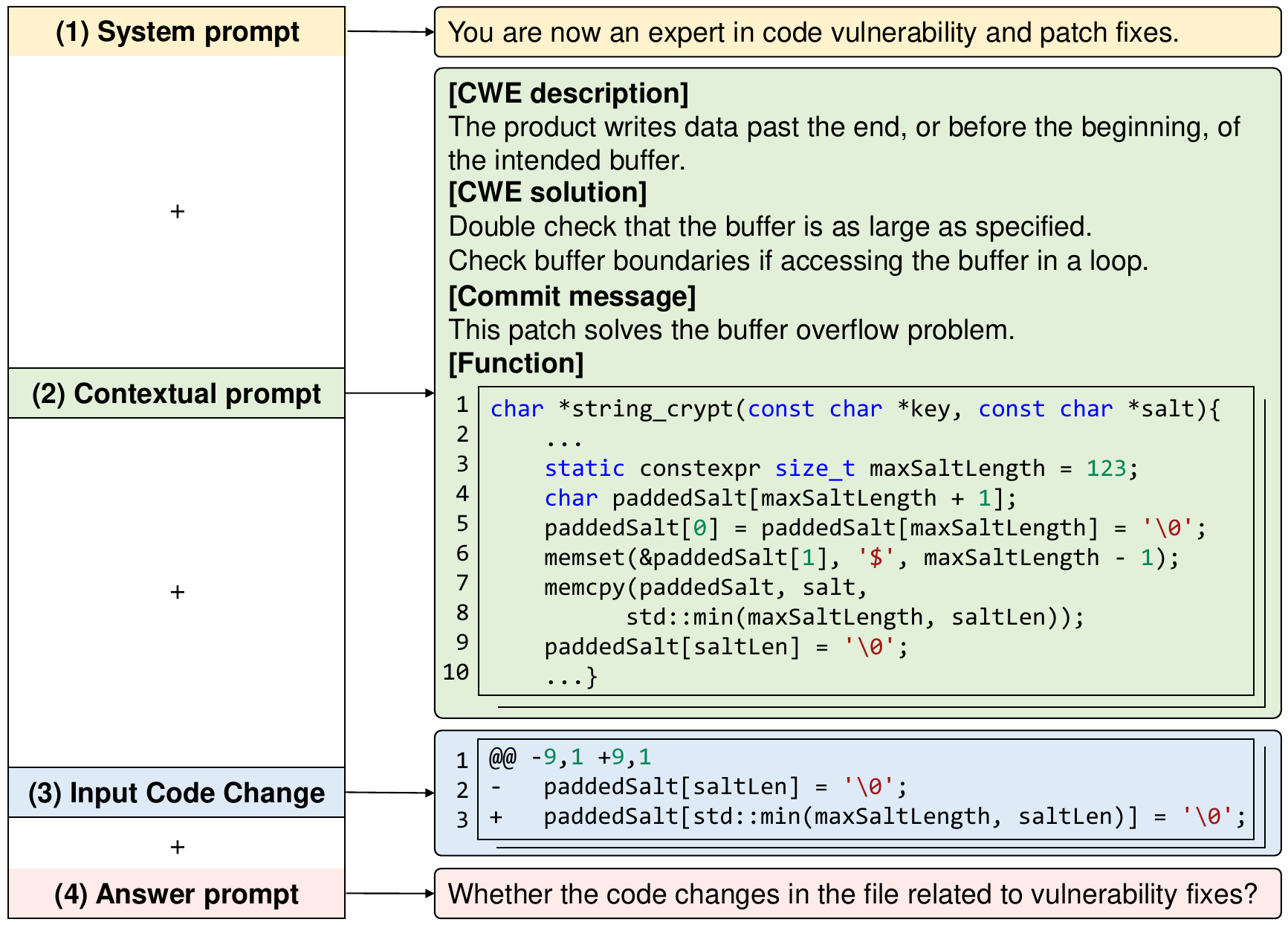}
    \caption{A sample prompt for LLMs to evaluate the relevance of code changes in one
    file to the vulnerability fixes.}
\label{prompt}
\end{figure}

To effectively leverage LLMs' contextual understanding capability,
we craft a task-specific prompt to evaluate the relevance of
code changes to the corresponding
vulnerability fixes, as depicted in Figure~\ref{prompt}. This prompt consists of several components:
\textbf{(1) System prompt:} 
The LLM acts as an expert, analyzing code vulnerabilities and their corresponding fixes.
\textbf{(2) Contextual prompt:} The prompt involves four types of contextual information. \textcircled{1} \underline{CWE description:} Based on the CWE-ID associated with each patch, we provide a brief vulnerability description. \textcircled{2} \underline{CWE solution:} We offer the recommended solutions according to the CWE-ID, which helps LLMs assess the alignment of code changes with the vulnerability solutions. \textcircled{3} \underline{Commit message:} The commit messages from patches aid LLMs in comprehending the purpose of the fixes. \textcircled{4} \underline{Function:} We utilize the Tree-sitter tool~\cite{Tree-sitter} to extract functions affected by the code changes in the file. This prompt contextualizes the code changes, aiding in a precise analysis.
\textbf{(3) Input Code Change: 
}
The specific code changes made in the file.
\textbf{(4) Answer prompt:} 
The LLMs evaluate the relevance of code changes in the file to vulnerability fixes and output
``YES'' or ``NO'' indicating whether
the code changes are vulnerability-fixing related. Given the prompt with detailed vulnerability and patch information, LLMs can effectively determine the relevance of code changes to vulnerability fixes based on the code context. 

\subsubsection{Static Analysis Tool Checking}

Besides LLMs, we also employ static analysis tools for vulnerability checking. Static analysis tools detect vulnerabilities in code by extracting source code models~\cite{static} and employing diverse vulnerability rules, ensuring high code coverage and low false-negative rates. We employ various static analysis tools to check for vulnerabilities in code changes, which integrate the vulnerability rules and expert knowledge from different tools. 

We choose
four well-established tools, including Cppcheck~\cite{Cppcheck}, Flawfinder~\cite{Flawfinder}, RATS~\cite{rats}, and Semgrep~\cite{Semgrep}, considering
their open-source availability and flexible configuration options, for achieving
effective vulnerability detection across multiple programming languages. For different languages, we employ different static analysis tools based on the applicability of the tools. Specifically, for C and C++ files, we use
all the four tools. For Python files, our analysis is conducted using RATS and Semgrep; while for Java files, we only use Semgrep.
For each file in a patch, we first employ static analysis tools to check whether the before-fixing file contains vulnerabilities. We then determine one file as vulnerability-fixing related only if the vulnerabilities detected in the before-fixing version\footnote{We aggregate the vulnerabilities detected by the multiple static analysis tools during the processing.} correspond to code changes in the patch.

\subsubsection{Joint Decision}


We combine the comprehension capabilities of LLMs and the domain knowledge of static analysis tools to identify \fileA in one patch. The files are labeled as related or not only if both LLMs and static analysis tools reach the same decisions.
Files with conflicting results from the LLMs and static analysis tools are excluded for the subsequent processing.

      
\subsection{Multi-granularity Dependency Extraction Module}

Existing vulnerability datasets mainly focus on function-level vulnerabilities, ignoring the rich context information.
To mitigate this challenge, we extract the inter-procedural call relationships of vulnerabilities throughout the repository and construct multiple-granularity information for each vulnerability patch, including information at repository level, file level, function level, and line level.


\subsubsection{Pilot Experience}


\begin{figure}[tb]
    \centering
    \includegraphics[width=.48\textwidth]{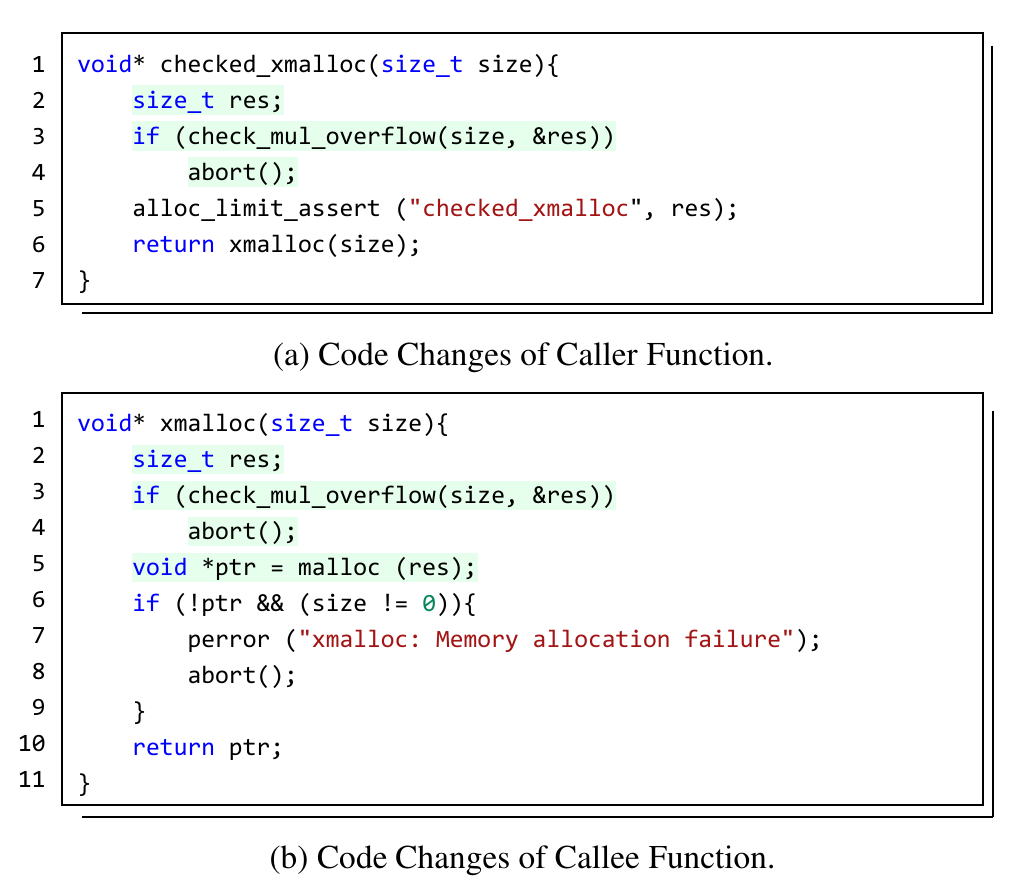}
    \caption{Two solutions to fix the inter-procedural vulnerability. The tokens highlighted in green indicate
    code changes related to vulnerability fixing.}
\label{example1}
\end{figure}


Figure~\ref{fig:example}(a) illustrates an example of an inter-procedural CWE-190 (Integer Overflow or Wraparound) vulnerability~\cite{CWE190}. 
Figure~\ref{example1}(a) and Figure~\ref{example1}(b) show two real-world solutions implemented across subsequent commits, which involve changing the caller function and adjusting the callee function, respectively. Specifically, Figure~\ref{example1}(a) demonstrates the first solution, where the function \texttt{check\_mul\_overflow()} is used to verify if the variable $size$ is overly large before the \texttt{xcalloc()} function is called (Lines 3-4). Figure~\ref{example1}(b) illustrates the second solution, which involves a check within the \texttt{xcalloc()} function itself to assess the size of $size$ (Lines 3-4).
Therefore, for the code snippets, whether function as callers
or callees, inter-procedural vulnerabilities can be introduced.
Note that code changes of Figure~\ref{example1}(a) do not cover the sixth line, which contains the inter-procedural vulnerability API, \texttt{xmalloc()}. Therefore, when capturing the inter-procedural vulnerability from a code snippet, we need to expand the search scope to its top-level functions.


\begin{algorithm}[tpb]
    \SetAlgoLined
    \footnotesize
    \SetKwInOut{Input}{Input}
    \SetKwInOut{Output}{Output}
    \SetKwInOut{Ensure}{Ensure}
    \SetKwFunction{Repository-level Dependency Extraction}{Repository-level Dependency Extraction}
    \SetKwProg{Fn}{Function}{:}{}
    \Input{$selLang$; // selected programming language \\
        $VulFile$; // all vulnerability-fixing related files in a project} 
    \Output{$callerTree$; // caller trees in the project \\
            $calleeTree$; // callee trees in the project}
    // initialize the $callerTree$ and $calleeTree$;\\
    $callerRoot \leftarrow \varnothing$; 
    $calleeRoot \leftarrow \varnothing$; \\
    \ForEach{$file \in VulFile$ }{
          { $project \leftarrow $ \texttt{getParentProject}($file$)}; \\
          {$allFile \leftarrow $ \texttt{getAllFile}$(project, selLang)$}; \\
          {$VulSnip \leftarrow $ \texttt{getSnip}$(file)$}; \\
        \ForEach{ $snip \in VulSnip$ }{
            {$callerRoot \leftarrow $ \texttt{getNearFunc}($snip, file$)}; \\
            {$calleeRoot \leftarrow $ \texttt{getOutFunc}($snip, file$)}; \\
            
            // adopt static analysis tools to extract caller and callee chains;\\
            {$callerTree$.add(\texttt{StaticToolExtractor}($callerRoot, allFile, 0$))}; \\
            {$calleeTree$.add(\texttt{StaticToolExtractor}($calleeRoot, allFile, 1$))}; 
        }
    }
    
\caption{Repository-level Dependency Extraction}
\label{algorithm1}
\end{algorithm}

\begin{algorithm}[t]
    \SetAlgoLined
    \footnotesize
    \SetKwInOut{Input}{Input}
    \SetKwInOut{Output}{Output}
    \SetKwInOut{Ensure}{Ensure}
    \SetKwFunction{Auxiliary Functions}{Auxiliary Functions}
    \SetKwProg{Fn}{Function}{:}{}
    // extract the root of the function callee tree; \\
    \Fn{\texttt{getOutFunc}($snip, file$)}{
        $ calleeFunc \leftarrow \varnothing $; \\
        // extract all the first layer function of $file$;\\
        $ outFunc \leftarrow $ \texttt{getTopLevelFunc}($file$); \\
        \ForEach{ $func \in outFunc$}{
            \If{$snip \cap func \neq \varnothing$}{
                {$ calleeFunc$.\texttt{add}($snip$); \\}
            }
        }
    }
    \Return{$calleeFunc$}\\
    // extract the root of the function caller tree;\\
    \Fn{\texttt{getNearFunc}($snip, file$)}{
    // expand the search scope to the top-level functions;\\
        $callerSnip \leftarrow$ \texttt{getOutFunc}($snip, file$);\\}
    \Return{\texttt{getAPI}($callerSnip, file$)}\\
    
\caption{Root Extraction for Caller and Callee Trees
}
\label{algorithm2}
\end{algorithm}

\subsubsection{Repository-level Dependency Extraction Algorithm}



We develop a dependency extraction algorithm comprising two components: \textit{Repository-level Dependency Extraction} (Algorithm~\ref{algorithm1}) and \textit{Root of Caller and Callee Tree Extraction} (Algorithm~\ref{algorithm2}). Algorithm~\ref{algorithm1} serves as the primary process for dependency extraction, while Algorithm~\ref{algorithm2} specifically addresses the extraction of the roots of function caller and callee trees.


Algorithm~\ref{algorithm1} presents our repository-level dependency extraction framework. It takes the selected programming language and all vulnerability-fixing related files 
in a project as input and produces two outputs: the function caller tree ($callerTree$) and the function callee tree ($calleeTree$). The algorithm works as follows: for each changed code snippet in a vulnerability-fixing related file, we initially identify the root functions for both caller and callee trees, named $callerRoot$ and $calleeRoot$, respectively (Lines 8-9). Subsequently, a specific static analysis tool is used to build the caller trees from $callerRoot$ and callee trees from $calleeRoot$ (Lines 11-12). For different programming languages, we use different tools: cflow~\cite{cflow} for C/C++, Java-all-call-graph~\cite{java-all-call-graph} for Java, and PyCG~\cite{PyCG} for Python, considering these tools are specifically designed for certain
programming languages.



Algorithm~\ref{algorithm2} is designed to identify the roots of function caller and callee trees. For the root of the function callee tree extraction, we identify all the top-level functions within the file (Line 5). Next, we retrieve the top-level functions that overlap with the provided code snippet (Lines 6-10). For the root of the function caller tree extraction, we first expand the scope of API extraction to the top-level functions containing the given code snippet (Line 15).
We then leverage the Tree-sitter tool~\cite{Tree-sitter} to extract all relevant APIs in the scope (Line 16).

\subsubsection{Multi-granularity Code Snippet}

To identify inter-procedural and intra-procedural vulnerabilities and facilitate the localization of vulnerabilities, we construct multiple-granularity information for each patch, including repository-level, file-level, function-level, and line-level. As shown in Table~\ref{tab:dataset_info_more}, each patch contains multi-granularity information.

\begin{table}[t]
    \caption{Multi-granularity information for each patch in \dataset, including file-level, function-level, file-level, and repository-level information.}
    \centering
    \setlength{\tabcolsep}{0.5mm}
    \renewcommand{\arraystretch}{1.0}
    \begin{tabular}{cl}
        \toprule
          \textbf{Features} & \textbf{Description} \\
        \midrule \midrule
        \multicolumn{2}{c}{\textbf{Line-level}} \\ \midrule 
            Line &  The content of the code line\\
            Line-Number &  The line number of the code line\\ \midrule \midrule
        \multicolumn{2}{c}{\textbf{Function-level}} \\ \midrule
             Function &  The content of the function\\
            Target &  The vulnerability label of the function\\ \midrule \midrule
        \multicolumn{2}{c}{\textbf{File-level}} \\ \midrule
        
          LLMs-Evaluate &  LLMs evaluation results\\ 
           Static-Check &  Static analysis tools checking results \\ 
          Target &  The vulnerability label of the file\\ 
           \midrule \midrule
        \multicolumn{2}{c}{\textbf{Repository-level}} \\ \midrule
        
             Inter-procedural Code & The content of the inter-procedural code\\
           Target  & The vulnerability label of the code snippet\\ 
        
        \bottomrule
    \end{tabular}
    \label{tab:dataset_info_more}
\end{table}

\textbf{Repository-level:}
For each vulnerability-fixing related file, we employ Algorithm~\ref{algorithm1} to extract the inter-procedural call relationships of vulnerabilities among the whole repository.

\textbf{File-level:}
We consider the vulnerability-fixing related files before and after code changes as vulnerable and non-vulnerable, respectively. For vulnerability-fixing unrelated files, we consider both the files before and after code changes
as non-vulnerable.

\textbf{Function-level:}
For each function affected by code changes, if the function is defined in \fileA, we consider the function before and after code changes as vulnerable and non-vulnerable, respectively; if the function is defined in \fileB, we consider both the function before and after code changes as non-vulnerable. For each function unaffected by code changes, we consider the function as non-vulnerable.

\textbf{Line-level:}
We extract the line changes and their line numbers from 
code changes in the patch and leverage these line changes as a precise detection target.

\subsection{Trace-based Filtering Module}

Patches may introduce vulnerabilities and become outdated, but the existing datasets do not distinguish outdated patches, posing a potential risk to data quality. In this module, we initially track the submission history of patches based on file paths and commit time. Through analyzing patches' historical information, we then recognize outdated patches by tracing their commit diffs.


\subsubsection{File Path Trace-based Filter}

We initially filter noise files according to the suffixes. For example, some files such as
description documentation (suffixed with \textit{.md, .rst}), data (suffixed with \textit{.json, .svg}), change logs (suffixed with \textit{.ChangeLog}), and output files (suffixed with \textit{.out}), are generally unrelated to functionality implementation. For the remained files, we create a dictionary associating file paths with their most recent submission dates in our collected vulnerability patches. We then review the submission date of each file, comparing it to the latest date recorded in the dictionary. If the file's submission date is not the most recent, we retain the file for the subsequent \moduleCb. Otherwise, we filter it out. As shown in Figure~\ref{fig:example}(c), the file \textit{ttm\_page\_alloc.c} still contains vulnerabilities after the first code changes. We create the entry in the dictionary based on the file's path \textit{``ttm\_page\_alloc.c''} and its latest submission date. According to the file's submission dates, we consider the earlier submitted file (commit id a66477b) as vulnerable.



\subsubsection{Commit Time Trace-based Filter}

We first retrieve each original patch's parent patch and child patch based on its commit time. 
We then assess whether the files changed within the parent patch and child patch overlap with those changed by the original patch. If there is an overlap within the retained files by the ~\moduleCa, we recognize the original patch as outdated. As illustrated in Figure~\ref{fig:example}(c), the original patch (commit id a66477b) and its child patch (commit id ac1e516) change the same file \textit{ttm\_page\_alloc.c} due to the incompleteness of the original patch. By comparing the overlapping
between the changed files in the original patch and its child patch, we can find the same changed file \textit{ttm\_page\_alloc.c}. Since the file is retained by the \moduleCa, we consider the original patch as outdated.



\section{Evaluation and Experimental Results}

\label{sec:evaluation}
In this section, we illustrate the advantages of \dataset and focus on the following four Research Questions (RQs):

\begin{enumerate}[label=\bfseries RQ\arabic*:,leftmargin=.5in]
    
    \item What are the advantages of \dataset compared to the existing vulnerability datasets?
    \item What is the quality of data labels in
    \dataset?
    \item To what extent do the
    selection of LLMs and prompt design in the \moduleA affect the label quality?
    \item How does \dataset perform in filtering outdated patches?

\end{enumerate}

\subsection{RQ1: Advantages of \dataset}

\definecolor{darkgreen}{rgb}{0,0.5,0}
\newcommand{\cross}{\textcolor{red}{\textbf{\XSolidBrush}}}
\newcommand{\tick}{\textcolor{darkgreen}{\Checkmark}}



\begin{table*}[htbp]
    \caption{Comparison between \dataset and six widely-used vulnerability detection datasets.}
    \setlength{\tabcolsep}{1.1mm}
    
    \centering
    \begin{tabular}{c|cccc|c|c|c|c}
    \toprule
               & \multicolumn{4}{c|}{\textbf{Granularity}} & \multirow{2}{*}{\textbf{CWE Types}} & \multirow{2}{*}{\textbf{Labeling Method}} & \textbf{Outdated Patch} & \multirow{2}{*}{\textbf{Additional Information}} \\ 
      \textbf{Baseline} & Line & Function & File & Repository & & & \textbf{Recognition} &   \\ \midrule
      BigVul~\cite{bigvul}   & \tick & \tick & \cross & \cross & 91 & Commit Code Diff & \cross & \tick \\ 
      D2A~\cite{D2A}   & \tick & \tick &  \cross & \cross & - & Static Analysis & \cross & \tick \\ 
      Devign~\cite{devign}  &  \cross & \tick &  \cross &  \cross  & - & Manually & \cross & \cross \\ 
      Reveal~\cite{reveal} &  \cross & \tick &  \cross &  \cross  & - & Commit Code Diff & \cross & \cross \\ 
      CrossVul~\cite{CrossVul} &  \cross & \cross &  \tick &  \cross  & 168 & Commit Code Diff & \cross & \tick \\ 
      DiverseVul~\cite{diversevul}  & \cross & \tick & \cross & \cross & 150 &  Commit Code Diff & \cross & \tick\\  \midrule
      \dataset & \tick & \tick &  \tick& \tick & 236 & LLMs + Static Analysis & \tick & \tick \\
    \bottomrule
    \end{tabular}

    \label{RQ1_compare}
\end{table*}

\begin{table}[tbp]
    \caption{Statistics of \dataset.}
    \centering
    \begin{tabular}{c|r|r|r|r}
    \toprule
      \textbf{Languages} & \textbf{CVE Entries} & \textbf{Patches} & \textbf{Files} & \textbf{Functions} \\ \midrule
      C & 3,565  & 4,010 &  7,515 &  212,790\\
      C++ & 631  & 689& 1,506 &  20,302\\ 
      Java & 786  & 888& 2,925 &  2,816\\ 
      Python & 1,152  & 1,310& 2,760 & 26,308 \\ 
      \midrule
      Total & 6,134  & 6,897& 14,706 &  262,216\\ 
    \bottomrule
    \end{tabular}
    \label{tab:data}
\end{table}

To answer RQ1, we compare \dataset with
six widely-used vulnerability detection datasets~\cite{bigvul, D2A, devign, reveal, CrossVul, diversevul} from several aspects, including the
granularity of vulnerabilities, number of CWE types, vulnerability labeling methods, outdated patches recognition, and additional information.
As shown in Table~\ref{RQ1_compare}, \dataset has the following advantages compared to current datasets:

\textbf{Multi-granularity information: }
Compared to other datasets that only contain vulnerabilities at the line-level, function-level, and file-level, \dataset contains more comprehensive granularities, including repository-level, file-level, function-level, and line-level vulnerabilities, which 
considers inter-procedural vulnerabilities and provides information that is more than a single patch. As shown in Table~\ref{tab:data}, \dataset comprises 14,706 files from 6,897 patches. Specifically, \dataset encompasses 212,790 functions in C, 20,302 in C++, 2,816 in Java, and 26,308 in Python, respectively.

\textbf{Extensive CWE coverage: }
 As shown in Table~\ref{RQ1_compare}, \dataset covers more CWE types than all other datasets. \dataset covers 149 CWE types in C, 105 in C++, 129 in Java, and 159 in Python, respectively. \dataset encompasses CWE types that extend across various programming languages since some CWE types are not language-specific. It indicates that \dataset provides more comprehensive data than existing benchmarks.



\textbf{Effective labeling methods: }
The previous work~\cite{quality} has identified the noisy data problem in the existing datasets by the current labeling method. In this paper, \dataset proposes the vulnerability untangling module for improving the vulnerability data quality. It involves the vulnerability rules and domain knowledge from static analysis tools and the strong contextual understanding capability from LLMs.

\textbf{Recognition of outdated patches: }
Current vulnerability datasets do not distinguish outdated patches. \dataset employs the \moduleC to recognize potential outdated patches. The \moduleC integrates \moduleCa and \moduleCb to provide
labels for outdated patches.

\textbf{Specific richness of additional information: } 
\dataset contains the richest additional information, including CVE descriptions, CVSS, and patch submission history illustrated in Table~\ref{tab:dataset_info}, and static analysis information illustrated in Table~\ref{tab:dataset_info_more}. Comprehensive information on vulnerabilities enables
developers and researchers to take effective measures for vulnerability detection.


 \begin{tcolorbox}
 \textbf{Answer to RQ1:} Compared to the existing datasets, 
 \dataset incorporates multi-granularity code snippets and the most extensive range of CWE types. It also employs an effective method for labeling vulnerability data and provides annotations for outdated patches, along with other rich additional information.
 \end{tcolorbox}
 
\subsection{RQ2: Label Quality of \dataset} 
\label{sec:rq2}
To answer the RQ2, we first conduct experiments
to assess the labeling method of \dataset (i.e., the \moduleA in the data collection framework). Then, we compare the label quality of \dataset with that of the previous datasets.

\textbf{Comparison with LLMs and static analysis tools:} 
We compare the label quality of \dataset with that obtained by LLMs and static analysis tools separately. The results are presented in Table~\ref{challenge2_manual}. Specifically, we randomly select 50 CVE cases for each programming language. We recruit three academic researchers as participants, and each of them possesses over five years of software vulnerability detection experience.
According to the vulnerability description and the commit message of the patch, the participants
manually label the code changes of the file in the patch as ``Yes'' or ``No'', corresponding to whether the code changes of the file are relevant to the vulnerability fix or not, respectively.
After assessing, participants reach agreements on 96\% for the cases. For the remaining discrepancies, participants negotiate and reach a consensus.
As shown in Table~\ref{challenge2_manual}, we observe that the proposed labeling method
achieves the highest accuracy of 85\% across the four programming languages on average, exceeding LLMs by 10\% and static analysis tools by 5.5\%.



\textbf{Comparison with the existing datasets:}
The previous work~\cite{quality} has shown that 20-71\% of vulnerability labels are inaccurate in the state-of-the-art OSS vulnerability datasets. Analyzing the results in Table~\ref{challenge2_manual}, we observe that \dataset shows the labeling accuracy at 85\%, 90\%, 85\%, and 80\% on C, C++, Java, and Python, respectively. Compared with the existing OSS vulnerability datasets as reported in~\cite{quality}, \dataset
achieves relatively higher accuracy
across the four programming languages, especially for the C++ programming language with an outstanding accuracy at 90\%.
Our observations indicate that \dataset's label quality is better than previous datasets
due to the integration of contextual understanding capability from LLMs and domain knowledge from static analysis tools. 

\begin{table}[tbp]
    \caption{Results of manual examination in 50 cases per programming language.}
    \setlength{\tabcolsep}{5mm}

    \centering
    \begin{tabular}{c|l|c}
    \toprule
      \textbf{Language}   & \multicolumn{1}{c|}{\textbf{Method}}  & \textbf{Accuracy} \\ \midrule
      \multirow{3}{*}{C}   &   LLMs & 70\%  \\ 
       & Static analysis tool &  82\%  \\ 
       &  \dataset &  \textbf{85\%}  \\ \midrule
       \multirow{3}{*}{C++}   &   LLMs & 80\%  \\ 
       &  Static analysis tool &  84\%  \\ 
       &  \dataset &  \textbf{90\%}  \\ \midrule
       \multirow{3}{*}{Java}   &   LLMs & 78\%  \\ 
       &  Static analysis tool &  78\%  \\ 
       &  \dataset  &  \textbf{85\%} \\ \midrule
       \multirow{3}{*}{Python}   &   LLMs & 72\%  \\ 
       &  Static analysis tool &  74\%  \\ 
       &  \dataset   & \textbf{80\%} \\ 
    \bottomrule
    \end{tabular}
    \label{challenge2_manual}
\end{table}

 \begin{tcolorbox}
 \textbf{Answer to RQ2:} After manual checking of a subset of labelled data, the label quality of \dataset 
 outperforms the existing datasets,
 achieving accuracy of 85\%, 90\%, 85\%, and 80\% on C, C++, Java, and Python, respectively. The proposed labeling method also performs better than LLMs and static analysis tools applied separately.
 \end{tcolorbox}
 
\subsection{RQ3: Influence of Different LLMs 
and Prompt Design on Label Quality}

To answer the RQ3, we conduct the experiments to analyze the impact of
LLM and prompt in the \moduleA on the label quality of \dataset. 
In this section, we randomly select 20 CVE cases, encompassing code changes that may or may not be associated with vulnerability fixes. We recruit the same three academic researchers who participate in the manual annotation of Section~\ref{sec:rq2} for labeling.
Participants manually label the code changes as ``Relevant'' and ``Irrelevant'', respectively. After assessing, participants reach agreements on 100\% of the 20 cases. 

\textbf{LLM selection: }
We present the nine investigated LLMs in Table~\ref{challenge1_llm}. ChatGLM~\cite{Chatglm} and Baichuan2~\cite{Baichuan} are open-source LLMs from THUDM and baichuan-ai, respectively, trained on Chinese and English corpora. Llama2~\cite{Llama2} boasts a larger training dataset compared to Llama~\cite{llama}. CodeLlama~\cite{Codellama} is trained and fine-tuned using code-related data based on Llama2. ChatGPT~\cite{ChatGPT} and GPT-4~\cite{GPT4} are commercial LLMs from OpenAI. Tongyi~\cite{TongYi} is developed by Alibaba Cloud, demonstrating excellent performance across multiple tasks. The experimental results are presented in
Table~\ref{challenge1_llm}. The results show that GPT-4 and Tongyi achieve the highest accuracy of 85\%, identifying all code changes related to vulnerability fixes and the majority of code changes unrelated to vulnerability fixes. Llama2, on the other hand, considers all code changes as relevant to vulnerability fixes and thus cannot effectively distinguish tangled patches. The other LLMs' accuracy does not exceed 80\%. Considering that Tongyi~\cite{TongYi} has the highest accuracy with free API accessibility, we choose this LLM for labeling.

\begin{table}[tbp]
    \caption{Performance of different LLMs evaluation in 20 CWE cases. The numbers split by ``/'' in the ``Relevant'' (or ``Irrelevant'') column represent the number of correct and incorrect responses in relevant (or irrelevant) code changes. The symbol ``-'' indicates that the corresponding statistic is unknown.}
    \setlength{\tabcolsep}{1.2mm}
    
    \centering
    \begin{tabular}{c|c|c|c|c}
    \toprule
      \textbf{Model}  & \textbf{Size} & \textbf{Relevant} & \textbf{Irrelevant} & \textbf{Accuracy} \\ \midrule
      Llama2-7b~\cite{Llama2}  & 7B & 4/0 & 0/16 & 20\% \\ 
      Llama2-13b~\cite{Llama2}  & 13B & 4/0 & 0/16 & 20\% \\ 
      Baichuan2~\cite{Baichuan} & 7B  & 1/3 & 9/7 & 50\% \\ 
      CodeLlama-7b~\cite{Codellama}  & 7B & 3/1 & 9/7 & 60\% \\
      CodeLlama-13b~\cite{Codellama}  & 13B & 3/1 & 10/6 & 65\% \\ 
      ChatGLM~\cite{Chatglm}  & 6B & 2/2 & 11/5 & 65\% \\ 
      ChatGPT~\cite{ChatGPT}  & 20B & 3/1 & 12/4 & 75\% \\ 
      GPT-4~\cite{GPT4}  & - & \textbf{4}/\textbf{0} & \textbf{13}/\textbf{3} & \textbf{85\%} \\ 
      Tongyi~\cite{TongYi}  & - & \textbf{4}/\textbf{0} & \textbf{13}/\textbf{3} & \textbf{85\%} \\
    \bottomrule
    \end{tabular}
    \label{challenge1_llm}
\end{table}

\begin{table}[tbp]
    \caption{Performance of different prompts evaluation in 20 CWE cases. The numbers split by ``/'' in the ``Relevant'' (or ``Irrelevant'') column represent the number of correct and incorrect responses in relevant (or irrelevant) code changes.}

    \centering
    \begin{tabular}{c|c|c|c}
    \toprule
      \textbf{Prompt}   &  \textbf{Relevant} & \textbf{Irrelevant} & \textbf{Accuracy} \\ \midrule
      w/o CWE Description  & 3/1 & 8/8 & 55\% \\ 
      w/o CWE Solution  & 3/1 & 11/5 & 70\% \\  
      w/o Commit Message & 3/1 & 13/3 & 80\% \\ 
      w/o Function   & 4/0 & 12/4 & 80\% \\ 
      \midrule
      \dataset   & \textbf{4}/\textbf{0} & \textbf{13}/\textbf{3} & \textbf{85\%} \\
    \bottomrule
    \end{tabular}
    \label{challenge1_prompt}
\end{table}

\textbf{Prompt design: }
The previous work~\cite{vul1} has demonstrated that the code-related task performance is largely influenced by the prompt. We also construct four variations, including prompt without CWE description (\ie w/o CWE Description), CWE solution (\ie w/o CWE Solution), commit message (\ie w/o Commit Message), and function of the code changes (\ie w/o Function). 
As illustrated in Table \ref{challenge1_prompt}, the prompt used by \dataset achieves the highest accuracy of 85\%. The four variations result in accuracy decreases of 30\%, 15\%, 5\%, and 5\%, respectively. The results indicate that the CWE description has the greatest impact on LLMs owing to its rich vulnerability information compared to the CWE solution and the function of the code changes. The relatively small impact of the commit message on the LLMs may be due to inaccurate and redundant information in the commit messages.

 \begin{tcolorbox}
 \textbf{Answer to RQ3:} 
 Among different LLMs, GPT-4 and Tongyi achieve the best accuracy of 85\% in labeling the relevance of code changes to
 vulnerability fixes. In prompt design, CWE description and CWE solution have a relatively considerable impact on the LLMs' decision. The proposed
 prompt achieves the highest accuracy of 85\%.
 \end{tcolorbox}

\subsection{RQ4: Performance of Filtering Outdated Patches in \dataset}


\begin{figure*}[h]
  \centering
    \includegraphics[width=\linewidth]{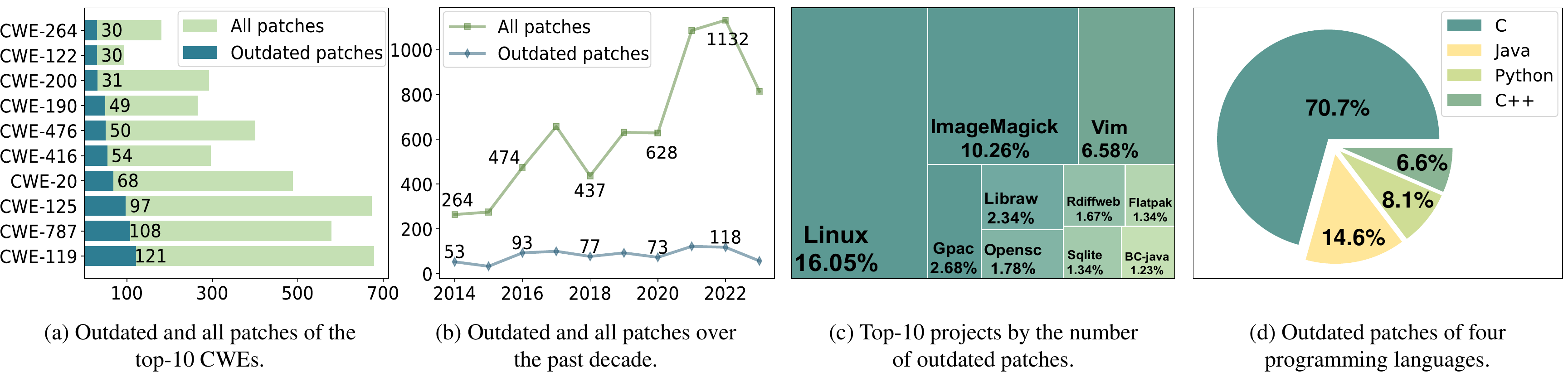}
  \caption{Outdated patches about CWEs, time, projects, and programming languages.}
  \label{challenge3}
\end{figure*}

To answer the RQ4, we present the statistics of recognized outdated patches by \dataset from different aspects, including CWEs, time, projects, and programming languages.

\textbf{CWEs:} 
Figure~\ref{challenge3}(a) shows the distribution of outdated patches and total patches of the Top-10 CWEs by the number of outdated patches. We observe that CWE-119 (Improper Restriction of Operations within the Bounds of a Memory Buffer)~\cite{CWE119}, CWE-787 (Out-of-bounds Write)~\cite{CWE787}, and CWE-125 (Out-of-bounds Read)~\cite{CWE125} contain the most outdated patches, including 121, 108, and 97 outdated patches, respectively.
These CWEs are associated with buffer operations, indicating that developers are more prone to introducing new vulnerabilities when modifying code snippets related to buffers, thereby resulting in outdated patches. 

\textbf{Time: }
Figure~\ref{challenge3}(b) depicts the distribution of outdated patches and total patches throughout the past decade. It 
reveals a consistent upward trend in the overall number of patches associated with OSS, ascending from 264 in 2014 to 1,132 in 2022. Concurrently, the count of outdated patches has seen a modest increase, progressing from 53 in 2014 to 118 in 2022. Notably, in 2023, both metrics exhibit a decline, which can be attributed to undisclosed vulnerabilities.
Moreover, the ratio of outdated patches to total patches demonstrates a notable decrease, plummeting from 20.07\% in 2014 to 10.42\% in 2022. This decline indicates a heightened emphasis by developers on enhancing the robustness of patches.

\textbf{Projects:}
Figure~\ref{challenge3}(c) presents the Top-10 projects ranked by the number of outdated patches. The proportion of outdated patches is highest in the Linux project, attaining
a rate of 16.05\%. Following closely are ImageMagick and Vim, exhibiting proportions of 10.26\% and 6.58\%, respectively. The heightened prevalence in Linux can be attributed to its expansive project scope and the vast number of files, making it susceptible to the introduction of new vulnerabilities during the submission of patches. Notably, ImageMagick and Vim manifest a considerable number of outdated patches despite their smaller project sizes. It may be due to delays of patch information maintained by the security vendors.

\textbf{Programming languages: }
Figure~\ref{challenge3}(d) illustrates the distribution of outdated patches categorized by programming languages. Among these languages, C constitutes the predominant share at 70.7\%, followed by Java at 14.6\%, Python at 8.1\%, and C++ at 6.6\%. 
The heightened occurrence within the C programming language is associated with
improper buffer operations related to CWE-119, CWE-787, and CWE-125. The vulnerabilities are
notably attributed to the incorporation of flexible arrays and the absence of built-in boundary-checking mechanisms.


 \begin{tcolorbox}
 \textbf{Answer to RQ4:} 
Within the identified outdated patches,
those associated with buffer operations comprise the majority. The quantity of outdated patches increases over time, yet their proportion in the total patches diminishes. Linux and C have the highest proportions of outdated patches by 16.05\% and 70.7\%, respectively, across
projects and programming languages.
 \end{tcolorbox}


\section{Discussion}

\label{sec:discussion}




\subsection{Data Application}
ReposVul is the first repository-level vulnerability dataset across
multiple programming languages. \dataset can be used for addressing a range of OSS vulnerability-related tasks. We advocate for the utilization of \dataset as a benchmark, promoting a
standardized and practical evaluation of model performance.

\textbf{Multi-granularity vulnerability detection:} \dataset covers
multi-granularity information, including repository-level, file-level, function-level, and line-level features. Researchers can leverage these features to detect inter-procedural and intra-procedural vulnerabilities. In addition, the experiments have demonstrated that \dataset outperforms the existing state-of-the-art datasets in label quality. It supports DL-based vulnerability detection methods for better training in the future.
 
\textbf{Patch management:} \dataset covers a rich set of patch information including the submission date, parent patches, and historical submission information of vulnerability patches. 
Researchers and practitioners can utilize this timely information to learn the patching process of existing software
vulnerabilities and identify outdated patches. 


\textbf{Vulnerability repair:} \dataset provides CVE descriptions, CWE solutions, CWE consequences, and CWE methods. This information aids developers to better understand the causes and solutions of vulnerabilities. Future research is expected to incorporate the rich contextual information for automatically repairing OSS vulnerabilities.

\subsection{Threats and Limitations}

One threat to validity comes from the collecting source platforms. During the collecting process, we collect \dataset from GitHub, Google Git, and bugs.chromium, which leads to missing projects that are hosted on other platforms in \dataset.

The second threat to validity is the programming language. We only extract repository-level dependency for four widely-used programming languages due to the language-specific features.
However, vulnerabilities also exist in other languages, such as JavaScript, Go, and PHP. 
We plan to extract repository-level dependency for more languages in future work.

Another validity to the threat comes from the collecting time, we only collect the CVEs from 2010 and the previous CVEs are not included in \dataset, which causes some vulnerabilities may be discovered and fixed in previous years.



\section{Related work}

\label{sec:related work}
\subsection{OSS Vulnerability Dataset}


The previous works utilize different methods for constructing OSS vulnerability datasets. It consists of manual checking-based, commit code diff-based, and static analysis tool generation.
Manual checking-based datasets~\cite{manual1, manual2, manual3, manual4} utilize test cases crafted artificially. For example, SARD ~\cite{SARD} includes samples from student-authored and industrial production.
Pradel et al.~\cite{Pradel} employ code transformation to convert non-vulnerable samples into vulnerable ones. 
However, the manual checking-based methods face limitations in labeling efficiency. Commit code diff-based datasets~\cite{bigvul, diversevul, reveal} gather patches from open-source repositories and extract vulnerability data from code changes in the patches.
CVEfixes~\cite{CVEfixes} fetches vulnerable entries from the National Vulnerability Database (NVD)~\cite{NVD} and corresponding fixes based on entries' reference links. CrossVul~\cite{CrossVul} generates an authentic dataset spanning 40 programming languages and 1,675 projects, while 
it only provides file-level source code. In addition, some datasets use static analysis tools~\cite{Infer, rats, Cppcheck} to label vulnerability data. For example, Russell et al.~\cite{Russell} integrate the results of multiple static analysis tools to determine whether vulnerabilities exist in code snippets.
D2A~\cite{D2A} collects vulnerabilities from six open-source projects and employs the Infer~\cite{Infer} to label data.

However, all these existing vulnerability datasets face challenges in label quality and lack inter-procedural vulnerability. 
In this paper, we propose the vulnerability untangling module for distinguishing vulnerability-fixing related code changes from tangled patches. We also propose the \moduleB for capturing the inter-procedural call relationships.

\subsection{OSS Vulnerability Detection}

OSS vulnerability detection is essential to identify security flaws and maintain software security. The existing OSS vulnerability detection methods consist of program analysis-based methods~\cite{fuzz1, fuzz2, static, static3} and learning-based vulnerability detection methods~\cite{vul2, vul3, vul5, vul6, wxc2, wxc3}.

The program analysis-based methods utilize expert knowledge in extracting features to identify vulnerabilities,
including data flow analysis~\cite{dataflow} and symbolic execution~\cite{symbolicexecution}. Data flow analysis~\cite{Flawfinder, rats, checkmarx, fortify, coverity} tracks the data flow along the execution paths of the program to obtain status information at program points, thus detecting vulnerabilities based on the security of program points. Symbolic execution~\cite{SAGE, S2E, mlsa} employs symbolic inputs instead of actual inputs and detects program vulnerabilities by determining whether symbolic expressions satisfy constraints. 

The learning-based methods can be classified into sequence-based~\cite{vul1, vul4, vul7, vul8} and graph-based methods~\cite{vul9, vul10, vul11, wxc1} due to the representation of the source code. Sequence-based methods transform code into token sequences. For example, VulDeePecker~\cite{VulDeePecker} utilizes code gadgets to represent programs and employs a Bidirectional Long Short-Term Memory (BiLSTM) network for training. $\mu$VulDeePecker~\cite{uVulDeePecker} combines three BiLSTM networks, enabling the detection of various types of vulnerabilities. Russell et al.~\cite{Russell} integrate convolutional neural networks and recurrent neural networks for feature extraction, utilizing a random forest classifier for capturing vulnerability patterns. Graph-based methods represent code as graphs and use graph neutral networks for software vulnerability detection. Qian et al.~\cite{qian} employ attributed control flow graphs to construct a vulnerability search engine. Devign~\cite{devign} adopts gated graph neutral networks to process multiple directed graphs generated from source code.

However, all these learning-based methods require a large amount of high-quality labeled samples to achieve good performance. In this paper, we construct a repository-level high-quality dataset named \dataset to promote the model training.

\section{Conclusion}

\label{sec:conclustion}
In this paper, we propose an automated data collection framework and construct the first repository-level vulnerability dataset named \dataset. 
Our framework consists of a \moduleA for identifying tangled patches, a \moduleB for extracting inter-procedural vulnerabilities, and a \moduleC for recognizing outdated patches. \dataset covers 6,134 CVE entries across 1,491 projects and four programming languages. After comprehensive data analysis and manual checking, \dataset proves to be high-quality and widely applicable compared with the existing vulnerability dataset.

Our source code, as well as \dataset, are available at \url{https://github.com/Eshe0922/ReposVul}.


\bibliographystyle{unsrt}
\bibliography{Citation}

\end{document}